\documentstyle[12pt]{article}
\def\ie{{\em i.e.}}
\def\eg{{\em e.g.}}

\def\beq{\begin{equation}}
\def\eeq{\end{equation}}

\catcode`\@=11 
\def\coeff#1#2{{\textstyle{#1\over #2}}}

\def\lsim{\mathrel{\mathpalette\@versim<}}
\def\gsim{\mathrel{\mathpalette\@versim>}}
\def\@versim#1#2{\vcenter{\offinterlineskip
    \ialign{$\m@th#1\hfil##\hfil$\crcr#2\crcr\sim\crcr } }}
\def\etal{{\em et. al.}}
\def\JL{J. L. Lopez}
\def\DVN{D. V. Nanopoulos}
\def\AZ{A. Zichichi}

\def\XW{X. Wang}

\def\t1{{\tilde 1}}

\def\GeV{\,{\rm GeV}}
\def\TeV{\,{\rm TeV}}

\def\to{\rightarrow}
\def\pb{\,{\rm pb}}
\def\fb{\,{\rm fb}}
\def\ipb{\,{\rm pb}^{-1}}
\def \met	{/\!\!\!\!E_{T}}
\def\NPB#1#2#3{Nucl. Phys. B {\bf#1} (19#2) #3}
\def\PLB#1#2#3{Phys. Lett. B {\bf#1} (19#2) #3}

\def\PRD#1#2#3{Phys. Rev. D {\bf#1} (19#2) #3}
\def\PRL#1#2#3{Phys. Rev. Lett. {\bf#1} (19#2) #3}

\def\MODA#1#2#3{Mod. Phys. Lett. A {\bf#1} (19#2) #3}
\def\IJMP#1#2#3{Int. J. Mod. Phys. A {\bf#1} (19#2) #3}
\def\TAMU#1{Texas A \& M University preprint CTP-TAMU-#1}

\begin{document}
\begin{flushright}
\baselineskip=12pt
{CERN-TH.7401/94}\\
{CTP-TAMU-27/94}\\
{ACT-09/94}\\
{hep-ph/9408345}
\end{flushright}

\begin{center}
\vglue 0.5cm
{\Large\bf Experimental consequences of one-parameter no-scale
supergravity models\\}
\vglue 1cm
{JORGE L. LOPEZ$^{(a),(b)}$, D. V. NANOPOULOS$^{(a),(b),(c)}$, and A.
ZICHICHI$^{(d)}$\\}
\vglue 0.4cm
{\em $^{(a)}$Center for Theoretical Physics, Department of Physics, Texas A\&M
University\\}
{\em College Station, TX 77843--4242, USA\\}
{\em $^{(b)}$Astroparticle Physics Group, Houston Advanced Research Center
(HARC)\\}
{\em Mitchell Campus, The Woodlands, TX 77381, USA\\}
{\em $^{(c)}$CERN Theory Division, 1211 Geneva 23, Switzerland\\}
{\em $^{(d)}$CERN, 1211 Geneva 23, Switzerland\\}
\baselineskip=12pt

\vglue 1cm
{\bf ABSTRACT}
\end{center}
{\rightskip=3pc
 \leftskip=3pc
\noindent
We consider the experimental predictions of two {\em one-parameter} no-scale
$SU(5)\times U(1)$ supergravity models with string-inspired moduli and
dilaton seeds of supersymmetry breaking. These predictions have been
considerably sharpened with the new information on the top-quark mass from
the Tevatron, and the actual measurement of the $B(b\to s\gamma)$ branching
ratio from CLEO. In particular, the sign of the Higgs mixing parameter $\mu$
is fixed. A more precise measurement of the top-quark mass above (below)
$\approx160\GeV$ would disfavor the dilaton (moduli) scenario. Similarly
a measurement of the lightest Higgs-boson mass above 90 GeV (below 100 GeV)
would disfavor the dilaton (moduli) scenario. At the Tevatron with $100\ipb$,
the reach into parameter space is significant only in the dilaton scenario
($m_{\chi^\pm_1}\lsim80\GeV$) via the trilepton and top-squark signals.
At LEPII the dilaton scenario could be probed up the kinematical limit
via chargino and top-squark pair production, and the discovery of the lightest
Higgs boson is guaranteed. In the moduli scenario only selectron pair
production looks promising. We also calculate the supersymmetric contribution
to the anomalous magnetic moment of the muon.}
\vspace{1cm}

\begin{flushleft}
\baselineskip=12pt
{CERN-TH.7401/94}\\
{CTP-TAMU-27/94}\\
{ACT-09/94}\\
August 1994
\end{flushleft}

\vfill\eject
\setcounter{page}{1}
\pagestyle{plain}
\baselineskip=14pt

\section{Introduction}
Experimental tests of supersymmetric models have become quite topical with
the advent of high-energy colliders such as the Tevatron and LEP, and their
planned or proposed upgrades. Since the expected scale of supersymmetry
is no lower than the electroweak scale, it is not surprising that
supersymmetric particles have yet to be found, even though they may be
``just around the corner". In working out the predictions for experimental
measurables, one must resort to sensible models of low-energy supersymmetry,
which in all generality are described by a large number of parameters. The
explicit or implicit choice of a theoretical ``framework" to reduce the size of
the parameter space is therefore mandatory. The typical framework consists
of a supergravity theory with minimal matter content and universal
soft-supersymmetry-breaking parameters, and radiative breaking of the
electroweak symmetry. Only four parameters are then needed to describe models
within this framework: $m_{1/2},m_0,A,\tan\beta$ \cite{reviews}.

Further reduction in the number of model parameters can be accomplished by
invoking the physics of superstrings, as the underlying theory behind the
effective supergravity model one chooses. However, the superstring framework is
likely not consistent with the Minimal Supersymmetric Standard Model (MSSM)
matter content, since unification of the gauge couplings  should not occur
until the string scale ($M_{\rm string}\sim10^{18}\GeV$) is reached. Moreover,
traditional GUT models are not easily obtained in string model-building.
Therefore we are led to consider a prototype GUST (Grand Unified Superstring
Theory) based on the gauge group $SU(5)\times U(1)$ \cite{revitalized} with
additional intermediate-scale particles to unify at the string scale
\cite{price+}. The motivations for $SU(5)\times U(1)$ model building have been
elaborated elsewhere \cite{EriceDec92}. Of particular relevance are the natural
suppression of dimension-five proton decay operators, the elegant
doublet-triplet splitting mechanism, and the novel see-saw mechanism.
Also, usual Yukawa coupling unification is not required in $SU(5)\times U(1)$.
Concrete string models based on the gauge group $SU(5)\times U(1)$ has also
been obtained and explored in detail \cite{Moscow}.

Within the string framework one can study ans\"atze for the
soft-supersymmetry-breaking parameters \cite{KL}. We consider such assumptions
which are also {\em universal}, and entail relations of the form:
$m_0=m_0(m_{1/2})$ and $A=A(m_{1/2}$).  We thus obtain a two-dimensional
parameter space ($m_{1/2},\tan\beta$). As a last step in the reduction process,
we study two specific scenarios in which the supersymmetry breaking parameter
associated with the Higgs mixing term $\mu$ is also determined (\ie,
$B=B(m_{1/2})$), \ie, `strict no-scale $SU(5)\times U(1)$ supergravity'.
These assumptions occur naturally in supersymmetry breaking
scenarios driven by the $F$-terms of the moduli or dilaton fields
\begin{itemize}
\item moduli scenario \cite{EKNI+II}:
\begin{equation}
m_0=0,\quad A=0,\quad B=0\,.
\label{moduli}
\end{equation}
\item dilaton scenario \cite{KL}:
\begin{equation}
m_0=\coeff{1}{\sqrt{3}}m_{1/2}\,,\quad A=-m_{1/2}\,,\quad
B=\coeff{2}{\sqrt{3}}m_{1/2}\,.
\label{dilaton}
\end{equation}
\end{itemize}
These two scenarios are {\em one-parameter} models, since we can trade
$B=B(m_{1/2})$ for $\tan\beta=\tan\beta(m_{1/2})$. They also have a dependence
on the top-quark mass, and the sign of the Higgs mixing parameter $\mu$.

The final parameter $m_{1/2}$, \ie, the scale of the supersymmetric spectrum,
can be determined dynamically in the no-scale supergravity framework \cite{LN}.
In this framework one starts with a supergravity theory with a flat direction
which leaves the gravitino mass undetermined at the classical level. This
theory also has two very healthy properties regarding the vacuum energy: (i) it
vanishes at tree-level \cite{Cremmer}, and (ii) it has no large one-loop
corrections (\ie, ${\rm Str}\,{\cal M}^2=0$) \cite{EKNI+II}. In this case the
vacuum energy is at most ${\cal O}(m^4_W)$. Minimization of the electroweak
effective potential with respect to the field corresponding to the flat
direction determines in principle the scale of supersymmetry breaking
\cite{Lahanas}. In this spirit we study the present one-parameter models
keeping in mind that the ultimate parameter will be determined eventually by
the no-scale mechanism in specific string models.

This paper is organized as follows. In section~\ref{sec:parspace} we explore
the constraints on the parameter space of these models and describe their
sparticle and Higgs-boson spectrum. In section~\ref{sec:bsg} we update the
calculation of $B(b\to s\gamma)$ and contrast it with the latest CLEO results.
In section~\ref{sec:exp} we study the experimental signals for these models at
the Tevatron (squark-gluino, trileptons, top-squarks, top-quark decays), LEPII
(Higgs bosons, charginos, selectrons, and top-squarks), and HERA (elastic
selectron-neutralino and chargino-sneutrino production). Finally in
section~\ref{sec:conclusions} we summarize our conclusions.

\section{Parameter space and spectrum}
\label{sec:parspace}
The one-dimensional parameter space of the models described above can be
represented in the $(m_{\chi^\pm_1},\tan\beta)$ plane by the relation
$\tan\beta=\tan\beta(m_{\chi^\pm_1})$. This exercise has been carried out for
the moduli and dilaton scenarios first in Refs.~\cite{LNZI} and \cite{LNZII}
respectively. The results depend on the value of $m_t$ and the sign of $\mu$.
In the moduli scenario one can show \cite{LNZI} that for $m_t\lsim130\GeV$
the condition in Eq.~(\ref{moduli}) ($B=0$) can only be satisfied for $\mu>0$,
whereas for $m_t\gsim135\GeV$ this condition requires $\mu<0$. In the
dilaton scenario the corresponding condition in Eq.~(\ref{dilaton})
($B=(2/\sqrt{3})m_{1/2}$) can only be satisfied for $\mu<0$ \cite{LNZII}.
The reason for these restrictions is that $\mu$ and $B$ are determined by
the radiative electroweak symmetry breaking constraint, and this depends on
$m_t$.

The above quoted values of $m_t$ are running masses which are related to
the experimentally observable pole masses via~\cite{GBGS}
\begin{equation}
m_t^{\rm pole}=m_t(m_t)\left[1+\coeff{4}{3}{\alpha_s(m_t)\over\pi}
+K_t\left({\alpha_s(m_t)\over\pi}\right)^2\right]
\label{pole}
\end{equation}
where
\begin{equation}
K_t=16.11-1.04\sum_{m_{q_i}<m_t}\left(1-{m_{q_i}\over m_t}\right)\approx11.
\end{equation}
Thus we obtain $m^{\rm pole}_t\approx 1.07 m_t$. Taking the recently announced
CDF measurement at face value, $m^{\rm pole}_t=174\pm17\GeV$ \cite{CDF}, we
can see that $m_t=130\GeV\leftrightarrow m^{\rm pole}_t=139\GeV$ is more than
$2\sigma$ too low. On the other hand, fits to all electroweak data
prefer a lower top-quark mass, when the Higgs-boson mass is restricted to be
light (as expected in a supersymmetric theory). The latest global fit gives
$m^{\rm pole}_t=162\pm9\GeV$ \cite{EFL}, and $m_t=130\GeV$ is again more than
$2\sigma$ too low. Thus we conclude that in the moduli scenario one must have
$\mu<0$, since $\mu>0$ can only occur for values of $m_t$ which are in gross
disagreement with present experimental data. Interestingly enough, both
one-parameter models are viable only for $\mu<0$. All sparticle and Higgs
boson masses and the calculations based on them have a dependence on $m_t$.
As discussed below, the $m_t$ dependence is small in the dilaton scenario,
but it can be significant in the moduli scenario. Unless otherwise stated, in
what follows we take $m_t=150\GeV\leftrightarrow
m^{\rm pole}_t=160\GeV$ as a representative value.

The calculated value of $\tan\beta$, as outlined above, is shown in
Fig.~\ref{PS-bsg}. The various symbols used to denote the points will be
discussed below.
\begin{itemize}
\item {\em Dilaton scenario}. The dependence
on $m_{\chi^\pm_1}$ is very mild: $\tan\beta\approx1.4$. The dependence on
$m_t$ is also mild; $m^{\rm pole}_t=160\GeV$ is shown in
Fig.~\ref{PS-bsg}. However, for $\tan\beta=1.4$ one must have
$m_t\lsim155\GeV\leftrightarrow m^{\rm pole}_t\lsim165\GeV$ in order to avoid a
Landau pole in the evolution of $\lambda_t$ up to the unification scale. This
upper limit on $m^{\rm pole}_t$ is well within all presently known limits on
$m_t$.
\item {\em Moduli scenario}. The results are rather $m_t$ dependent
($m^{\rm pole}_t=160,170,180\GeV$ are shown in Fig.~\ref{PS-bsg}),
with $\tan\beta$ decreasing with increasing values of $m_t$:
\begin{eqnarray}
\tan\beta&\approx&17.7-23.1\quad {\rm for}\quad m^{\rm
pole}_t=150\GeV,\nonumber\\
\tan\beta&\approx&12.4-18.4\quad {\rm for}\quad m^{\rm
pole}_t=160\GeV,\nonumber\\
\tan\beta&\approx&8.25-13.1\quad {\rm for}\quad m^{\rm
pole}_t=170\GeV,\nonumber\\
\tan\beta&\approx&5.53-8.53\quad {\rm for}\quad m^{\rm
pole}_t=180\GeV.\nonumber
\end{eqnarray}
The range of $\tan\beta$ values indicates the monotonic increase with
$m_{\chi^\pm_1}$.
\end{itemize}

For a chosen value of $m_t$ we can then calculate the sparticle and
Higgs boson spectrum as a function of $m_{\chi^\pm_1}$. After this is
done, the current experimental lower bounds on the sparticle and Higgs-boson
masses (most importantly $m_h\gsim64\GeV$ and $m_{\chi^\pm_1}\gsim45\GeV$)
are enforced and the actual allowed parameter space results. The neutralino
and chargino masses are shown in Fig.~\ref{NC}, the slepton and Higgs-boson
masses, and the value of $\mu$ in Fig.~\ref{sleph}, and the gluino and squark
masses in Fig.~\ref{sqg}. One can see that most of the masses are nearly linear
functions of $m_{\chi^\pm_1}$, although the slope depends on the scenario.
Common result are the relations:
\begin{eqnarray}
m_{\chi^\pm_1}&\approx&m_{\chi^0_2} \approx2m_{\chi^0_1},\\
m_{\chi^0_3}&\approx&m_{\chi^0_3}\approx m_{\chi^\pm_2}\approx|\mu|,\\
m_{\tilde g}&\approx& 3.3 m_{\chi^\pm_1}+90\GeV.
\end{eqnarray}
Another common feature is the little variability of $m_h$ with
$m_{\chi^\pm_1}$:
\begin{eqnarray}
m_h&\approx&100-115\GeV,\qquad{\rm moduli\ scenario};\label{higgs_moduli}\\
m_h&\approx&64-88\GeV,\qquad{\rm dilaton\ scenario}.\label{higgs_dilaton}
\end{eqnarray}
In fact, the two Higgs-mass ranges do not overlap and thus a measurement
of $m_h$ would discriminate between the two models. Note also that the three
Higgs bosons beyond the lightest one are very close in mass (especially in the
dilaton scenario, see Fig.~\ref{sleph}) and not light:
$m_{\rm H^+}\gsim160\,(400)\GeV$ in the moduli (dilaton) scenario.

In the moduli scenario the stau mass eigenstates are significantly split
from the corresponding selectron (and smuon) masses (see Fig.~\ref{sleph}).
This is not the case for the dilaton scenario where they are largely
degenerate. These facts are easily understood in terms of the value of $m_0$ in
the two scenarios. Regarding the gluino and squark masses (see Fig.~\ref{sqg}),
the ``average" squark mass (\ie, $m_{\tilde q}=(m_{\tilde u_L}+m_{\tilde
u_R}+m_{\tilde d_L}+m_{\tilde d_R})/4$, with about $\pm2\%$ split around the
average) is very close to the gluino mass
\begin{eqnarray}
m_{\tilde q}&\approx&0.97m_{\tilde g}\gsim255\GeV,\qquad{\rm moduli\
scenario};\\
m_{\tilde q}&\approx&1.01m_{\tilde g}\gsim260\GeV,\qquad{\rm dilaton\
scenario};
\end{eqnarray}
and both are higher than the present Tevatron sensitivity.
We note in passing that the potentially large difference between the running
gluino mass $m_{\tilde g}$ and the pole gluino mass $m^{\rm pole}_{\tilde g}$
\cite{gluino}
\begin{equation}
m^{\rm pole}_{\tilde g}=m_{\tilde g}\left[1+{\alpha_s\over\pi}
\left(3+\coeff{1}{4}\sum_{\tilde q}\log{m_{\tilde q}\over m_{\tilde
g}}\right)\right]
\end{equation}
is naturally suppressed in these models because of the near degeneracy of
gluino and squark masses. We find $m^{\rm pole}_{\tilde g}\approx1.1m_{\tilde
g}$.

The situation is quite different for the third-generation squark masses,
which are significantly split from $m_{\tilde q}$, except for $\tilde b_2$:
$m_{\tilde b_2}\approx m_{\tilde q}$. The most striking departure is that of
the lightest top-squark $\tilde t_1$. The implications of a light top-squark
in the dilaton scenario have been recently discussed in Ref.~\cite{stop},
and will be re-emphasized below. We should add that in this scenario the small
value of $\tan\beta$ ($\approx1.4$) and the light top-squarks would appear to
imply very small tree-level ($\propto\cos^22\beta$) and one-loop
($\propto\ln(m_{\tilde t_1}m_{\tilde t_2}/m^2_t)$) contributions to $m^2_h$.
In practice, the one-loop contribution to $m_h$ turns out to be sizeable
enough because of the often-neglected top-squark mixing effect which adds
a large positive term $\sim(m^2_{\tilde t_1}-m^2_{\tilde t_2}){1\over2}
{\sin^22\theta_t\over m^2_t}\ln(m^2_{\tilde t_1}/m^2_{\tilde t_2})$ to the
usual piece \cite{ERZ}.

As noted above, in the moduli scenario the allowed values of $\tan\beta$ depend
strongly on $m^{\rm pole}_t$. The calculated value of $\mu$ also depends on
$m_t$: $\mu(m_t)\propto m_t$ to good approximation in the range of interest.
The lightest Higgs-boson mass also depends on $m_t$: the upper limit
in Eq.~(\ref{higgs_moduli}) increases to 120 (125) for $m^{\rm
pole}_t=170\,(180)\GeV$. The squarks and sleptons are only slightly affected
via their (small) $\tan\beta$ dependence. Even the lightest top-squark is
not affected by more than $\pm2\%$.

We have also calculated the relic abundance of the lightest neutralino
$\Omega_\chi h^2_0$ following the methods of Ref.~\cite{LNYdm}. The results
are shown in Fig.~\ref{DM}. We get $\Omega_\chi h^2_0\lsim0.25\,(0.9)$ in
the moduli (dilaton) scenario. These results are automatically consistent
with cosmological expectations (\ie, $\Omega_\chi h^2_0<1$). The structure
on the curves (especially in the dilaton scenario) corresponds to $s$-channel
$h$ and $Z$ poles in the annihilation cross section. The dotted line in the
moduli scenario reflects the $m_t$ dependence ($m^{\rm pole}=180\GeV$ for
this line) via the different $\tan\beta$ and $\mu$ values.

\section{$b\to s\gamma$}
\label{sec:bsg}
There are several indirect experimental constraints which can be applied
to $SU(5)\times U(1)$ supergravity models. For the case of two-parameter
models of this kind these constraints have been discussed in
Ref.~\cite{Easpects}. It turns out that the only one of relevance for the
one-parameter models is that from $B(b\to s\gamma)$. An analysis of this
constraint in a variety of supergravity models has recently been performed in
Ref.~\cite{LargeTanB}. Here we update this analysis for the one-parameter
models in light of the most recent CLEO experimental result \cite{newCLEO}
\begin{equation}
B(b\to s\gamma)=(1-4)\times10^{-4},\qquad{\rm at\ 95\%C.L.}
\label{bsg}
\end{equation}

In Fig.~\ref{bsgB} we show the calculated value of $B(b\to s\gamma)$ in both
scenarios (for $m^{\rm pole}_t=160\GeV$). The latest CLEO limits are
indicated by the solid lines, with the arrows pointing into the allowed
region. The Standard Model result is also shown. As explained in
Ref.~\cite{LargeTanB}, there is significant theoretical
uncertainty on the value of $B(b\to s\gamma)$, mostly from next-to-leading
order QCD corrections. We have roughly quantified this uncertainty by
using a leading order calculation but allowing the renormalization scale to
vary between $m_b/2$ and $2m_b$. This variation gives the dotted lines
above and below the solid lines in Fig.~\ref{bsgB}. The same procedure is
used to estimate the Standard Model uncertainty, which shows that the data
agree well with the Standard Model. In fact, the theoretical uncertainty in
the Standard Model prediction is {\em larger} than the present $1\sigma$
experimental uncertainty \cite{newCLEO}. In the moduli
scenario there is further uncertainty because the value of $m_t$ affects
the calculated value of $\tan\beta$. From Fig.~\ref{PS-bsg} we see that larger
values of $m^{\rm pole}_t$ decrease $\tan\beta$, and this leads to larger
values of $B(b\to s\gamma)$: the dash-dot line represents the result
($\mu=m_b$) for $m^{\rm pole}_t=180\GeV$.

In the one-parameter models, we consider points in parameter space to be
``excluded" if their interval of uncertainty does not overlap with that in
Eq.~(\ref{bsg}); these are denoted by crosses (`$\times$') in
Fig.~\ref{PS-bsg}. In the moduli scenario, this constraint requires
$m_{\chi^\pm_1}\gsim120\GeV$ for $m^{\rm pole}_t=160\GeV$, but only
$m_{\chi^\pm_1}\gsim75\GeV$ for $m^{\rm pole}_t=170\GeV$, and there is no
constraint for $m^{\rm pole}_t=180\GeV$. In the dilaton scenario, because the
allowed values of $\tan\beta$ are small, the constraint is rather
mild, only requiring $m_{\chi^\pm_1}\gsim50\GeV$. Strictier constraints
can be obtained by allowing less experimental uncertainty (\eg, $1\sigma$)
or less theoretical uncertainty, both of which are unwise things to do.
We have also identified ``preferred" points whose interval of uncertainty
overlaps with the corresponding Standard Model interval; these are denoted
by diamonds (`$\diamond$') in Fig.~\ref{PS-bsg}. From this vantage point,
the dilaton scenario or the moduli scenario with somewhat heavy top-quark look
quite promising.

\section{Experimental predictions}
\label{sec:exp}
We now discuss the experimental signatures of these one-parameter models
at the Tevatron, LEPII, and HERA. For this analysis we consider only the points
still allowed by the $b\to s\gamma$ constraint, \ie, those denoted by dots and
diamonds in Fig.~\ref{PS-bsg}.
\subsection{Tevatron}
We consider the present-day $\sqrt{s}=1.8\TeV$ Tevatron with an estimated
integrated luminosity of $\sim100\ipb$ at the end of the on-going Run IB.
Three supersymmetric signals could be observed: trileptons from
chargino-neutralino production, large missing energy from squark-gluino
production, and soft dileptons from top-squark production. All three signals
could be observable in the dilaton scenario; only
the squark-gluino signal may be observable in the moduli scenario.
\begin{itemize}
\item {\em Neutralinos and charginos}. These are produced in the reaction
$p\bar p\to\chi^\pm_1\chi^0_2X$ \cite{EHNS}, and when required to decay
leptonically yield a trilepton signal \cite{trileptons}. The cross section is
basically a monotonically decreasing function of $m_{\chi^\pm_1}$, whereas the
leptonic (and hadronic) branching fractions (given in Fig.~\ref{br}) are
greatly model dependent and vary as a function of the single parameter. Our
calculations have been performed as outlined in Ref.~\cite{LNWZ}. In the moduli
scenario there is an enhancement of the chargino leptonic branching fraction
because of the presence of light sleptons (\eg, $\chi^\pm_1\to
\ell^\pm\tilde\nu$). However, this gain is undone by the suppressed neutralino
leptonic branching fraction also because of the light sleptons (\ie,
$\chi^0_2\to \nu\tilde\nu$). In contrast, in the dilaton scenario both
branching fractions are comparable. However, in this case for
$m_{\chi^\pm_1}\gsim165\GeV$ the spoiler mode $\chi^0_2\to\chi^0_1 h$ opens
up and the neutralino leptonic branching fraction becomes negligible.

The trilepton rates are given in Fig.~\ref{trileptons}, where we indicate by a
dashed line the present CDF upper limit obtained with $\sim20\ipb$ of data
\cite{Kato}. By the end of the on-going Run~IB the integrated luminosity is
estimated at $\sim100\ipb$ per detector. If no events are observed, one could
estimate an increase in sensitivity by a factor of 4 (assuming 80\% efficiency
in recording data) per experiment. Combining both experiments the sensitivity
would be even higher (say $\sim6$ times better). In Fig.~\ref{trileptons} we
show the estimated sensitivity range as the area between the dotted lines. In
the dilaton scenario we estimate the reach at $m_{\chi^\pm_1}\lsim(80-90)\GeV$.
On the other hand, in the moduli scenario the rates are small, but there may
be a small observable window for $m_{\chi^\pm_1}\approx100\GeV$. However,
because of the constraints from $b\to s\gamma$, such values of $m_{\chi^\pm_1}$
are allowed only for $m^{\rm pole}_t\gsim165\GeV$.

\item {\em Gluino and squarks}. Since in these models we obtain $m_{\tilde
q}\approx m_{\tilde g}$, the multi-jet missing-energy signal is enhanced.
In both scenarios we also obtain a lower bound of $\sim260\GeV$ which makes
this signal almost kinematically inaccessible. Indeed, the reach with $100\ipb$
is estimated at $m_{\tilde q}\approx m_{\tilde g}\lsim300\GeV\leftrightarrow
m_{\chi^\pm_1}\lsim60\GeV$ \cite{James}.

\item {\em Top-squarks}.
Direct $\tilde t_1$ pair production at the Tevatron (via the dilepton mode)
has been shown recently \cite{BST} to be sensitive to $m_{\tilde
t_1}\lsim100\GeV$ by the end of Run~IB, provided the chargino leptonic
branching fraction is taken to be $\sim20\%$. In the dilaton scenario $\tilde
t_1$ can be rather light ($m_{\tilde t_1}\gsim70\GeV$) and the chargino
branching fractions are $\sim40\%$ (see Fig.~\ref{br}). In Ref.~\cite{BST},
with a ``bigness" $B=|p_T(\ell^+)|+|p_T(\ell'^-)|+|\met|$ cut of $B<100\GeV$,
the $t\bar t$ (with $m_t=170\GeV$) and $W^+W^-$ backgrounds are estimated
at $14\fb$ and $10\fb$ respectively. With $100\ipb$, a $5\sigma$ signal above
this background requires $(\sigma B)_{\rm dileptons}\gsim75\fb$. From Fig.~10
in Ref.~\cite{BST} it appears then that $m_{\tilde t_1}\lsim130\GeV$ could be
probed in this case of enhanced branching fractions. In the moduli scenario the
top-squarks are too heavy to be detectable ($m_{\tilde t_1}\gsim160\GeV$).

\begin{table}[t]
\caption{Cross sections at the Tevatron (in pb) for $p\bar p\to\tilde
t_1\bar{\tilde t_1}X$ [29] and $p\bar p\to t\bar tX$~[30]. All masses in
GeV.}
\label{Xsections}
\begin{center}
\begin{tabular}{|c|c|c|c|c|c|}\hline
$m_{\tilde t_1}$&70&80&90&100&112\\ \hline
$\sigma(\tilde t_1\bar{\tilde t_1})$&60&30&15&8&4 \\ \hline
\end{tabular}
\begin{tabular}{|c|c|c|c|c|}\hline
$m_t$&120&140&160&180\\ \hline
$\sigma(t\bar t)$&39&17&8&4 \\ \hline
\end{tabular}
\end{center}
\hrule
\end{table}

\item{\em Soft dileptons}. In the dilaton scenario, if $\tilde t_1$ is light
enough, events may already be present in the existing data sample. The cross
section for pair-production of the lightest top-squarks $\sigma(\tilde
t_1\bar{\tilde t_1})$ depends solely on $m_{\tilde t_1}$ \cite{BDGGT} and is
given for a sampling of values in Table~\ref{Xsections}.
Since in the dilaton scenario $m_{\tilde t_1}>m_{\chi^\pm_1}+m_b$ (see
Fig.~\ref{sqg}), one gets $B(\tilde t_1\to b\chi^\pm_1)=1$ (neglecting the
small one-loop $\tilde t_1\to c\chi^0_1$ mode \cite{HK}).\footnote{This
unsuppressed two-body decay mode of $\tilde t_1$ implies that scalar
stoponium will decay too soon to be observable \cite{stoponium}.} The charginos
then decay leptonically or hadronically with branching fractions shown in
Fig.~\ref{br}, \ie, $B(\chi^\pm_1\to \ell\nu_\ell\chi^0_1)\approx0.4$
($\ell=e+\mu$) for $m_{\chi^\pm_1}\lsim65\GeV\leftrightarrow m_{\tilde
t_1}\lsim100\GeV$. The most promising signature for light top-squark detection
is through the dilepton mode \cite{BST}. The ratio of stop-dileptons to
top-dileptons is \cite{stop}
\begin{equation}
{N^{\tilde t_1\bar{\tilde t_1}}_{2\ell}\over N^{t\bar t}_{2\ell}}
\approx 3.2{\sigma(\tilde t_1\bar{\tilde t_1})\over \sigma(t\bar t)}\ .
\label{ratio}
\end{equation}
This ratio indicates that for sufficiently light top-squarks there
may be a significant number of dilepton events of non--top-quark origin,
if the experimental acceptances are tuned accordingly.

Perhaps the most important distinction between top-dileptons and
stop-dileptons is their $p_T$ distribution: the (harder) top-dileptons come
from the two-body decay of the $W$ boson, whereas the (softer) stop-dileptons
come from the (usually) three-body decay of the chargino with masses (in
this case) below $m_W$. Therefore, the top-dilepton data sample is essentially
distinct from the stop-dilepton sample. Such distinction is well quantified by
the ``bigness" ($B$) parameter mentioned above.

\item {\em Top-quark branching fractions}. In the case of a light top-squark,
the channel $t\to \tilde t_1\chi^0_1$ may be kinematically accessible. In
the dilaton scenario for $m^{\rm pole}_t=160\GeV$ this is the case for
$m_{\tilde t_1}\lsim115\GeV$. The calculated values of $B(t\to \tilde
t_1\chi^0_1)$ and $B(t\to bW)$ are shown in Fig.~\ref{tbr}. One can see that
if $\tilde t_1$ is light enough, one would expect up to $(0.9)^2\approx20\%$
reduction in the number of observed top events relative to the Standard Model
prediction. However, this discrepancy would not be observable until a sizeable
top-quark sample is collected. For a $2\sigma$ effect one would need to measure
the $t\bar t$ cross section to $10\%$ accuracy, which requires $\sim100$
background-subtracted top events. This event sample will not be available
before the Main Injector era.

\end{itemize}

\subsection{LEPII}
At present it is uncertain what the LEPII beam energy may ultimately be. It is
expected that LEPII will turn on in 1996 at $\sqrt{s}\approx180\GeV$, while the
highest possible center-of-mass energy is estimated at $\sqrt{s}=240\GeV$.
The precise value of $\sqrt{s}$ has two main effects: it determines the
kinematical reach for pair-produced particles (such as charginos and
selectrons), and it determines the reach in Higgs-boson masses. The latter
is of more relevance since for sufficiently high values of
$\sqrt{s}(<240\GeV)$, it may be possible to cover {\em all} of the parameter
space of these models. For definiteness, unless otherwise stated, in what
follows we will set $\sqrt{s}=200\GeV$. We consider four signals: Higgs bosons,
charginos, charged sleptons, and top-squarks. The calculations of the first
three signals have been performed as described in Ref.~\cite{LNPWZ}.

\begin{itemize}
\item{\em Higgs bosons}. These are produced via $e^+e^-\to Zh$, with
$h\to b\bar b$ and $b$-tagging to reduce the background. The cross section
for this process differs from the corresponding Standard Model cross section
in two ways: by the factor $\sin^2(\alpha-\beta)$, and by the ratio
$f=B(h\to b\bar b)/B(H\to b\bar b)_{\rm SM}$. In the models under
consideration, $\sin^2(\alpha-\beta)$ is very close to 1, according to a
decoupling phenomenon induced by the radiative electroweak breaking mechanism
\cite{LNPWZh}. In Fig.~\ref{hbr} we show the $h\to b\bar b$ branching fraction
which shows that $f$ is usually close to 1, except when the supersymmetric
decay mode $h\to\chi^0_1\chi^0_1$ is open. This channel is open for
$m_h>2m_{\chi^0_1}\approx m_{\chi^\pm_1}$, \ie, only for the lightest values of
$m_h$, since $m_h$ grows little with $m_{\chi^\pm_1}$, as Fig.~\ref{hbr} shows.

The effective cross sections $\sigma(e^+e^-\to Zh)\times f$ are shown in
Fig.~\ref{higgs}. The deviations of the curves from monotonically decreasing
functions of $m_h$, which coincide with the Standard Model prediction, are due
to the $h\to\chi^0_1\chi^0_1$ erosion of the preferred $h\to b\bar b$ mode.
These deviations could be used to differentiate between the Standard Model
Higgs boson and the supersymmetric Higgs bosons considered here.

The LEPII sensitivity for Higgs-boson detection is estimated
at $0.2\pb$ for a $3\sigma$ effect in $500\ipb$ of data \cite{Sopczak}. In the
moduli scenario this cross section level is reached for
$m_h\approx106\,(114)\GeV$ for $\sqrt{s}=200\,(210)\GeV$. From
Eq.~(\ref{higgs_moduli}) we see that LEPII would need to run at
$\sqrt{s}\approx210\GeV$ to cover the whole parameter space (for $m^{\rm
pole}_t=160\GeV$). On the other hand, in the dilaton scenario for
$\sqrt{s}=200\GeV$ one obtains $\sigma(e^+e^-\to Zh)\times f>0.57\pb$, which
has a $5\sigma$ significance for ${\cal L}=170\ipb$. (For $\sqrt{s}=180\GeV$ we
obtain $\sigma(e^+e^-\to Zh)\times f>0.22\pb$, \ie, also observable with
sufficiently large ${\cal L}$.) Therefore, LEPII should be able to cover {\em
all} of the parameter space in the dilaton scenario via the Higgs signal.
In Fig.~\ref{hbr} we also show a detail of the relation between $m_h$ and
$m_{\chi^\pm_1}$ which shows that a lower bound on $m_h$ would immediately
translate into a lower bound on the chargino mass, and in turn into a lower
bound on all sparticle masses.

\item{\em Charginos}. The preferred signal is the ``mixed" decay ($1\ell+2j$)
in pair-produced charginos. The chargino branching fractions in Fig.~\ref{br}
indicate that this mode is healthy in the dilaton scenario, but rather
suppressed in the moduli scenario. This is revealed in Fig.~\ref{mixed} where
we show the ``mixed" cross sections in both scenarios. With an estimated
$5\sigma$ sensitivity of $0.12\pb$ (for $500\ipb$) \cite{Easpects}, one should
be able to reach up to $m_{\chi^\pm_1}\lsim96\GeV$ in the dilaton scenario. The
reach should decrease by $\sim10\GeV$ for $\sqrt{s}=180\GeV$. In the moduli
scenario we obtain $(\sigma B)_{\rm mixed}<0.03$, \ie, an unobservable signal.
However, the much larger dilepton mode may lead to an observable signal in this
case if the $W^+W^-$ background could somehow be dealt with.

\item{\em Sleptons}. At LEPII only in the moduli scenario are the sleptons
kinematically accessible (see Fig.~\ref{sleph}). The processes of interest are
$e^+e^-\to\tilde e_R\tilde e_R+\tilde e_R\tilde e_L$ and
$e^+e^-\to\tilde\mu_R\tilde\mu_R$, with the further decays $\tilde e_{R,L}\to
e\chi^0_1$ and $\tilde\mu_R\to\mu\chi^0_1$ with near 100\% branching fractions
\cite{LNPWZ}. The selectron cross section, shown in Fig.~\ref{slepton}, is
large: $\sigma(e^+e^-\to\tilde e\tilde e)>1\pb$ for $m_{\tilde e_R}<80\GeV$.
(Note the kink on the curve when the $\tilde e_R\tilde e_L$ channel closes.)
The $5\sigma$ sensitivity for ${\cal L}=100\,(500)\ipb$ is estimated at
$0.47\,(0.21)\pb$ \cite{Easpects}. This sensitivity level is reached for
$m_{\tilde e_R}\approx84\,(90)\GeV$ and corresponds to
$m_{\chi^\pm_1}\approx92\,(103)\GeV$ (see Fig.~\ref{sleph}). Thus, the {\em
indirect} reach for chargino masses is larger than the direct one (via the
``mixed" mode). The smuon cross section is much smaller (see
Fig.~\ref{slepton}) and so are the corresponding reaches in smuon
($m_{\tilde\mu_R}\approx70\,(82)\GeV$) and chargino masses. Again, these
reaches should decrease by $\sim10\GeV$ for $\sqrt{s}=180\GeV$.

\item {\em Top-squarks}. These are kinematically accessible only in the
dilaton scenario. Moreover, since $m_{\tilde t_1}\lsim100\GeV$ for
$m_{\chi^\pm_1}\lsim65\GeV$, the reach into the parameter space is not very
significant, but a new lower bound on $m_{\tilde t_1}$ could be obtained.
We consider the process
\begin{equation}
e^+e^-\to \tilde t_1\tilde t_1\to (b\chi^+_1)(\bar b\chi^-_1)
\end{equation}
with $B(\tilde t_1\to b\chi^\pm_1)=1$. The cross section $\sigma(e^+e^-\to
\tilde t_1\tilde t_1)$ proceeds through $s$-channel photon and $Z$ exchanges.
In the case of $Z$-exchange, the coupling $Z\tilde t_1\tilde t_1$ is
proportional to $\cos^2\theta_t-{4\over3}\sin^2\theta_W$ and vanishes for
$\cos^2\theta_t\approx0.31$. In the dilaton scenario, for $m_{\tilde
t_1}\lsim100\GeV$ we find $\cos^2\theta_t\approx0.60$ and this cancellation
does not occur. The cross section is shown in Fig.~\ref{stop1}, and has been
calculated including initial state radiation and QCD corrections, as described
in Ref.~\cite{DH}. Depending on the chargino decays, one can have three
signatures: $2b+2\ell$, $2b+1\ell+2j$, $2b+4j$, all with the same branching
fraction of $\approx(0.4)^2$ (see Fig.~\ref{br}). The traditional $W^+W^-$
background is not relevant (unless one $\ell$ is lost and there is no
$b$-tagging), and probably the channel with the least number of jets (\ie,
$2b+2\ell$) is preferable. Assuming a suitably cut background, three signal
events (of any of the three signatures) would be observed for $\sigma(e^+e^-\to
\tilde t_1\tilde t_1)\gsim0.19\,(0.04)\pb$ with ${\cal L}=100\,(500)\ipb$. From
Fig.~\ref{stop1} this sensitivity requirement implies a reach of $m_{\tilde
t_1}\approx85\,(95)\GeV$.
\end{itemize}

\subsection{HERA}
The weakly interacting sparticles may be detectable at HERA if they are light
enough and if HERA accumulates and integrated luminosity ${\cal O}(100\ipb)$.
So far HERA has accumulated a few $\ipb$ of data and it is expected that
eventually it will be producing $25-30\ipb$ per year. The supersymmetric
signals in $SU(5)\times U(1)$ supergravity have been studied in
Ref.~\cite{hera}, where it was shown that the elastic scattering signal, \ie,
when the proton remains intact, is the most promising one. The deep-inelastic
signal has smaller rates and is plagued with large backgrounds. The reactions
of interest are $e^- p\to \tilde e^-_{L,R}\chi^0_{1,2}p$ and $e^-p\to
\tilde\nu_e\chi^-_1p$. The total elastic supersymmetric signal is shown in
Fig.~\ref{hera} versus the chargino mass. The dashed line represents the limit
of sensitivity with ${\cal L}=200\ipb$ which will yield five ``supersymmetric"
events. The signal is very small in the dilaton scenario, but may be observable
in the moduli scenario. However, considering the timetable for the LEPII and
HERA programs, it is quite likely that LEPII would explore all of the HERA
accessible parameter space before HERA does. This outlook may change if new
developments in the HERA program would give priority to the search for the
right-handed selectron ($\tilde e_R$) which could be rather light in the moduli
scenario. At HERA one could also produce top-squark pairs, if they are light
enough, as possible in the dilaton scenario. However, the cross section in
this case is smaller than $0.01\pb$ for $m_{\tilde t_1}>70\GeV$ and decreases
very quickly with increasing values of $m_{\tilde t_1}$ \cite{Kon}.

\section{Conclusions}
\label{sec:conclusions}
We have explored the experimental consequences of well motivated one-parameter
no-scale $SU(5)\times U(1)$ supergravity in the moduli and dilaton scenarios.
Such models are highly predictive and therefore falsifiable through the
many correlations among the experimental observables. In fact, the recent
information on the top-quark mass has in effect ruled out half of the parameter
space in the moduli scenario, selecting the sign of mu to be negative.
Interestingly, this is also the required sign in the dilaton scenario. The
top-quark mass dependence is particularly important in these models in other
ways as well. In the moduli scenario the calculated values of $\tan\beta$
depend strongly on $m_t$ (see Fig.~\ref{PS-bsg}), and thus so do the calculated
values of $B(b\to s\gamma)$, which become quite restrictive for $m^{\rm
pole}_t\lsim160\GeV$. On the other hand, in the dilaton scenario values of
$m^{\rm pole}_t\gsim165\GeV$ are not allowed. Therefore, future more precise
determinations of the top-quark mass are likely to disfavor one the scenarios
and support the other. A similar (and partially related) dichotomy is present
in the Higgs-boson masses, which are below 90 GeV in the dilaton scenario and
above 100 GeV in the moduli scenario.

We concentrated on the experimental signals at present-day facilities: the
Tevatron with the expected integrated luminosity at the end of the ongoing
run, the forthcoming LEPII upgrade, and HERA. At the Tevatron the traditional
squark-gluino signal is enhanced (since $m_{\tilde q}\approx m_{\tilde g}$)
but the possible reach into parameter space is small since $m_{\tilde q}\approx
m_{\tilde g}\gsim260\GeV$ is required. The trilepton signal is more promising,
although only in the dilaton scenario, where a reach of
$m_{\chi^\pm_1}\approx(80-90)\GeV$ is expected. In this same scenario light
top-squarks could be detected for $m_{\tilde t_1}\lsim130\GeV\leftrightarrow
m_{\chi^\pm_1}\lsim80\GeV$. So through different channels the reach into the
parameter space should be similar. In the moduli scenario the reach into
parameter space is not promising.

At LEPII the Higgs boson should be readily detectable in the dilaton scenario
(for $\sqrt{s}>180\GeV$), in effect covering the whole parameter space of the
model. In fact, even an improved lower bound on $m_h$ will constrain the
parameter space immediately by requiring a lower bound on the chargino mass.
In the moduli scenario Higgs detection requires $\sqrt{s}\gsim200-210\GeV$.
In both scenarios, for sufficiently low values of $m_h$, the supersymmetric
channel $h\to\chi^0_1\chi^0_1$ is open and decreases the usual $b\bar b$
yield in a way which could be used to differentiate between the Standard Model
Higgs boson and the supersymmetric Higgs bosons considered here.
Charginos should be readily detectable (via the ``mixed" mode) almost
up to the kinematical limit ($\sqrt{s}/2$) in the dilaton scenario, but will be
hard to detect in the moduli scenario. On the other hand, selectrons should
be detectable up to near the kinematical limit in the moduli scenario
(corresponding to charginos slightly over the direct kinematical limit), and be
kinematically inaccessible in the dilaton scenario. Top-squarks in the dilaton
scenario should also be detectable up to near the kinematical limit, although
this corresponds to much lighter chargino masses than in the other detection
modes.

We conclude that at the Tevatron and LEPII the dilaton scenario is
significantly more accessible than the moduli scenario is:
\begin{center}
\begin{tabular}{|l|c|c|c||c|c|c|c|}\hline
&\multicolumn{3}{c||}{Tevatron}&\multicolumn{4}{c|}{LEPII}\\ \hline
&$\tilde q-\tilde g$&$\chi^\pm_1$&$\tilde t_1$&$h$&$\chi^\pm_1$&$\tilde
e$&$\tilde t_1$\\ \hline
moduli&$\surd$&$\times$&$\times$&$\surd$&$\times$&$\surd$&$\times$\\
dilaton&$\surd$&$\surd$&$\surd$&$\surd$&$\surd$&$\times$&$\surd$\\ \hline
\end{tabular}
\end{center}
Moreover, in the dilaton scenario LEPII is basically assured the discovery of
the Higgs boson. Also, LEPII has the possibility of increasing its reach in
both scenarios by increasing its center-of-mass energy.

There is one set of experimental observables which we have not discussed here,
namely the one-loop corrections to the LEP observables and their dependence
on the supersymmetric parameters. In the context of $SU(5)\times U(1)$
supergravity these observables have been discussed in
Ref.~\cite{epsilons,Easpects}, where it was concluded that as long as $m_t^{\rm
pole}\lsim170\GeV$ there are no constraints on the model parameters at the 90\%
C.L. One of these observables, namely the ratio $R_b=\Gamma(Z\to b\bar
b)/\Gamma(Z\to {\rm hadrons})$ has been measured more precisely during the past
year \cite{Schaile} and its value still remains more than $2\sigma$ above the
Standard Model prediction, for not too small values of the top-quark mass.
Recently in Ref.~\cite{KKW} it has been argued that supersymmetry could provide
a better fit to this observable should the chargino and the lightest top-squark
be both light. These conditions could be satisfied in the dilaton scenario
discussed above, although an explicit calculation of this observable is
required to be certain.

As a final experimental consequence of these models, we have calculated the
supersymmetric contribution to the anomalous magnetic moment of the muon (as
described in Ref.~\cite{g-2}), which is shown in Fig.~\ref{g-2}. The arrow
points into the presently experimentally allowed region. The upcoming
Brookhaven E821 experiment (1996) \cite{Roberts} aims at an experimental
accuracy of $0.4\times10^{-9}$, which is much smaller than the moduli scenario
prediction. This indirect experimental test is likely to be much more stringent
than any of the direct tests discussed above.
{
To close, we re-iterate that these one-parameter no-scale supergravity models
would become ``no-parameter" models once the no-scale mechanism is implemented
in specific string models, thereby determining the value of the ultimate
parameter.

\section*{Acknowledgments}
J.L. would like to thank James White for useful discussions. This work has been
supported in part by DOE grant DE-FG05-91-ER-40633.
}

\newpage

\begin{figure}[p]
\vspace{6.0in}
\includegraphics{PS-bsg.ps}
\caption{The one-dimensional parameter space of strict no-scale $SU(5)\times
U(1)$ supergravity -- moduli and dilaton scenarios. In the moduli scenario
results are rather $m_t$ dependent ($m^{\rm pole}_t=160,170,180\GeV$ are
shown). In the dilaton scenario $m^{\rm pole}_t=160\GeV$ is taken and
$m^{\rm pole}_t<165\GeV$ is required. Points excluded by $B(b\to
s\gamma)$ are denoted by crosses (`$\times$'), those consistent with the
Standard Model prediction are denoted by diamonds (`$\diamond$'), and the rest
are denoted by dots (`$\cdot$').}
\label{PS-bsg}
\end{figure}
\clearpage

\begin{figure}[p]
\vspace{6.0in}
\includegraphics{NC.ps}
\caption{The chargino and neutralino masses (in GeV) versus the lightest
chargino mass in strict no-scale $SU(5)\times U(1)$ supergravity -- moduli
and dilaton scenarios. The following relations hold: $m_{\chi^0_2}\approx
m_{\chi^\pm_1}\approx2m_{\chi^0_1}$, $m_{\chi^0_{3,4}}\approx
m_{\chi^\pm_2}\approx|\mu|$.}
\label{NC}
\end{figure}
\clearpage

\begin{figure}[p]
\vspace{6.0in}
\includegraphics{sleph.ps}
\caption{The slepton ($\tilde e_{L,R},\tilde\tau_{1,2},\tilde\nu$) and
Higgs-boson masses ($h,A,H,H^+$) as a function of the chargino mass in strict
no-scale $SU(5)\times U(1)$ supergravity -- moduli and dilaton scenarios. Also
shown is the calculated value of the Higgs mixing parameter~$\mu$.}
\label{sleph}
\end{figure}
\clearpage

\begin{figure}[p]
\vspace{6.0in}
\includegraphics{sqg.ps}
\caption{The first- and second-generation average squark mass ($\tilde q$),
the gluino mass ($\tilde g$, dashed lines), the sbottom masses ($\tilde
b_{1,2}$, $m_{\tilde b_2}\approx m_{\tilde q}$), and the stop masses
($\tilde t_{1,2}$) as a function of the chargino mass in strict no-scale
$SU(5)\times U(1)$ supergravity -- moduli and dilaton scenarios. Note
$m_{\tilde q}\approx m_{\tilde g}$ and that $\tilde t_1$ can be quite light in
the dilaton scenario ($m_{\tilde t_1}>67\,{\rm GeV}$).}
\label{sqg}
\end{figure}
\clearpage

\begin{figure}[p]
\vspace{6.0in}
\includegraphics{DM.ps}
\caption{The calculated value of the relic density of the
lightest neutralino ($\Omega_\chi h^2_0$) as a function of the chargino mass in
strict no-scale $SU(5)\times U(1)$ supergravity -- moduli and dilaton scenarios
for $m^{\rm pole}_t=160\GeV$. The dotted line in the moduli scenario
corresponds to $m^{\rm pole}_t=180\GeV$. Note that $\Omega_\chi h^2_0$ is
sizeable but within cosmological limits.}
\label{DM}
\end{figure}
\clearpage

\begin{figure}[p]
\vspace{6.0in}
\includegraphics{bsgB.ps}
\caption{The branching fraction $B(b\to s\gamma)$ as a function of the chargino
mass in strict no-scale $SU(5)\times U(1)$ supergravity -- moduli and dilaton
scenarios for $m^{\rm pole}=160\GeV$. The dotted lines above and below the
solid line indicate the estimated theoretical error in the prediction. The
dashed lines delimit the Standard Model prediction. The arrows point into the
currently experimentally allowed region. The dot-dash line in the moduli
scenario corresponds to $m^{\rm pole}_t=180\GeV$ (central value).}
\label{bsgB}
\end{figure}
\clearpage

\begin{figure}[p]
\vspace{6.0in}
\includegraphics{br.ps}
\caption{The chargino and neutralino leptonic and hadronic branching fractions
as a function of the chargino mass in strict no-scale $SU(5)\times U(1)$
supergravity -- moduli and dilaton scenarios. The sudden drops in the
neutralino branching fractions correspond to the opening of the ``spoiler"
mode $\chi^0_2\to\chi^0_1 h$.}
\label{br}
\end{figure}
\clearpage

\begin{figure}[p]
\vspace{6.0in}
\includegraphics{trileptons.ps}
\vspace{-1.5in}
\caption{The rate for trilepton production at the Tevatron as a function of the
chargino mass in strict no-scale $SU(5)\times U(1)$ supergravity -- moduli and
dilaton scenarios. The present CDF upper bound is indicated by the dashed
line, and the estimated reach at the end of Run IB is bounded by the dotted
lines.}
\label{trileptons}
\end{figure}
\clearpage

\begin{figure}[p]
\vspace{6.0in}
\includegraphics{tbr.ps}
\vspace{-1.5in}
\caption{Top-quark branching fractions for $m^{\rm pole}_t=160\,{\rm GeV}$
in strict no-scale $SU(5)\times U(1)$ supergravity -- dilaton scenario.}
\label{tbr}
\end{figure}
\clearpage

\begin{figure}[p]
\vspace{6.0in}
\includegraphics{hbr.ps}
\vspace{-1.5in}
\caption{The lightest Higgs-boson mass as a function of the chargino mass
in strict no-scale $SU(5)\times U(1)$ supergravity -- moduli and dilaton
scenarios. Note that the predicted mass ranges do not overlap and that a lower
bound on $m_h$ translates into a lower bound on the chargino mass.
Also shown are the Higgs-boson branching fractions into $b\bar b$ and
$\chi^0_1\chi^0_1$.}
\label{hbr}
\end{figure}
\clearpage

\begin{figure}[p]
\vspace{6.0in}
\includegraphics{higgs.ps}
\vspace{-1.5in}
\caption{The effective cross section $\sigma(e^+e^-\to Zh)\times f$
[$f=B(h\to b\bar b)/B(H\to b\bar b)_{\rm SM}$] for Higgs boson production at
LEPII (for the indicated center-of-mass energies) as a function of the
Higgs-boson mass in strict no-scale $SU(5)\times U(1)$ supergravity -- moduli
and dilaton scenarios. The dashed line indicates the estimated experimental
sensitivity. Note that the two scenario predictions do not overlap. The
deviations of the curves from monotonically decreasing functions of $m_h$
(which coincide with the Standard Model prediction) are due to the
$h\to\chi^0_1\chi^0_1$ erosion of the preferred $h\to b\bar b$ mode, and could
be used to differentiate between the Standard Model Higgs boson and the
supersymmetric
Higgs boson considered here.}
\label{higgs}
\end{figure}
\clearpage

\begin{figure}[p]
\vspace{6.0in}
\includegraphics{mixed.ps}
\vspace{-1.5in}
\caption{The chargino pair-production cross section into the ``mixed" mode
($1\ell+2j$) at LEPII as a function of the chargino mass in strict no-scale
$SU(5)\times U(1)$ supergravity -- moduli and dilaton scenarios.  The dashed
line indicates the estimated experimental sensitivity. Note that the two
scenario predictions do not overlap.}
\label{mixed}
\end{figure}
\clearpage

\begin{figure}[p]
\vspace{6.0in}
\includegraphics{slepton.ps}
\vspace{-1.5in}
\caption{The selectron ($\tilde e_R\tilde e_R+\tilde e_R\tilde e_L$)
and smuon ($\tilde\mu_R\tilde\mu_R$) pair-production cross sections at LEPII as
functions of the slepton mass in strict no-scale $SU(5)\times U(1)$
supergravity -- moduli scenario. The dashed lines delimit the estimated
experimental sensitivity. In the $\tilde e\tilde e$ case, for $m_{\tilde
e_R}>80\GeV$ the $\tilde e_R\tilde e_L$ channel is closed and thus the kink
on the curve.}
\label{slepton}
\end{figure}
\clearpage

\begin{figure}[p]
\vspace{6.0in}
\includegraphics{stop1.ps}
\vspace{-1.5in}
\caption{The lightest top-squark pair-production cross section at LEPII as a
function of the top-squark mass in strict no-scale $SU(5)\times U(1)$
supergravity -- dilaton scenario. The higher (lower) dashed line indicates the
limit of three $2b+2\ell$, $2b+1\ell+2j$, or $2b+4j$ events in ${\cal
L}=100\,(500)\,{\rm pb}^{-1}$ of data.}
\label{stop1}
\end{figure}
\clearpage

\begin{figure}[p]
\vspace{6.0in}
\includegraphics{hera.ps}
\vspace{-1.5in}
\caption{The total elastic supersymmetric cross section (including
selectron-neutralino and sneutrino-chargino production) at HERA
as a function of the chargino mass in strict no-scale $SU(5)\times U(1)$
supergravity -- moduli and dilaton scenarios.  The dashed line indicates the
sensitivity limit to observe five events in $200\,{\rm pb}^{-1}$ of data.}
\label{hera}
\end{figure}
\clearpage

\begin{figure}[p]
\vspace{6.0in}
\includegraphics{g-2.ps}
\caption{The supersymmetric contribution to the anomalous magnetic moment of
the muon ($a_\mu^{\rm susy}$) as a function of the chargino mass in strict
no-scale $SU(5)\times U(1)$ supergravity -- moduli and dilaton scenarios. The
arrow points into the currently experimentally allowed region.  In the moduli
scenario, results are rather $m_t$ dependent ($m^{\rm pole}_t=160,180\GeV$ are
shown). In the dilaton scenario $m^{\rm pole}_t=160\GeV$ is taken and
$m^{\rm pole}_t<165\GeV$ is required.}
\label{g-2}
\end{figure}
\clearpage

\end{document}